\documentclass[prb,twocolumn]{revtex4}%

\usepackage{graphicx}
\usepackage{amsmath}
\usepackage{ulem}
\usepackage{color}

\newcommand{\be}{\begin{equation}}
\newcommand{\ee}{\end{equation}}
\newcommand{\bea}{\begin{eqnarray*}}
\newcommand{\eea}{\end{eqnarray*}}
\newcommand{\bean}{\begin{eqnarray}}
\newcommand{\eean}{\end{eqnarray}}

\begin{document}

\draft
\title
{\bf Thermoelectric effects of quantum dot arrays embedded in
nanowires}

\author{Yen-Chun Tseng$^{1}$, David M.-T. Kuo$^{1,2,\dagger}$, Yia-Chung Chang$^{3,4,*}$ and Chia-Wei Tsai$^{1}$}
\address{$^{1}$Department of Electrical Engineering and $^{2}$Department of Physics, National Central
University, Chungli, 32001 Taiwan}

\address{$^{3}$Research Center for Applied Science, Academic Sinica, Taipei, 11529, Taiwan and
$^{4}$Department of Physics, National Cheng-Kung University, Tainan,
70101, Taiwan}

\date{\today}

\begin{abstract}
The thermoelectric properties of quantum dot arrays (QDAs) embedded
in nanowires connected to electrodes are studied theoretically in
the Coulomb blockade regime. A Hurbbard-Anderson model is used to
simulate the electronic contribution to thermoelectric properties of
a QDA junction system. The electrical conductance, Seebeck
coefficient, and electron thermal conductance are calculated by both
the Keldysh Green function method and the mean-field approach. The
phonon thermal conductivities are calculated by using the equation
of phonon radiative transfer method. In the Coulomb blockade regime
the electron thermal conductance is much smaller than the phonon
thermal conductance. Therefore, the optimal figure of merit ($ZT$)
can be enhanced by increasing thermal power and decreasing phonon
thermal conductance simultaneously. We found that it is possible to
obtain $ZT$ value of InGaAs/GaAs QDAs embedded in nanowires larger
than one at room temperature.

%\vspace{1cm} Keywords: Quantum dot array; Coulomb blockade; Seebeck
%coefficient; Figure of merit
\end{abstract}

\maketitle
\section{Introduction}

%\textbf{1. Introduction}

Recently, many efforts have been devoted to seeking efficient
thermoelectric (TE) materials with the figure of merit ($ZT$) larger
than 3, because there are potential applications of solid state
thermal devices such as coolers and power generators, which can
replace conventional compressor-based refrigerators and fossil fuel
generators to reduce $CO_2$ emission [1-7]. Nevertheless, the
optimization of TE materials is extremely difficult, since
$ZT=S^2G_eT/\kappa$ depends on the electrical conductance ($G_e$),
Seebeck coefficient (S),and thermal conductance ($\kappa$). $T$ is
the equilibrium temperature. These physical quantities are usually
related to one another. Mechanisms leading to the enhancement of
power factor ($PF=S^2G_e$) would also enhance the thermal
conductance. Consequently, it is difficult to obtain $ZT$ above one
in conventional bulk materials.[1]

Impressive $ZT$ values for quantum dot array (QDA) embedded in
nanowires have been experimentally demonstrated.[8] The power factor
and thermal conductance become independent thermoelectric variables
under the condition $\kappa_e/\kappa_{La} \ll 1$, where $\kappa_e$
and $\kappa_{La}$ denote, respectively, the electron thermal
conductance and lattice thermal conductance.[8] In the Coulomb
blockade regime, electron transport process is seriously suppressed
by the electron Coulomb interactions, therefore $\kappa_e$ as well
as $G_e$ are reduced significantly.[9] Under the condition
$\kappa_e/\kappa_{La} \ll 1$, one can increase the power factor and
decrease the phonon thermal conductance simultaneously to optimize
$ZT$.[9]

Thermoelectric properties of quantum dots (QDs) embedded in a matrix
connected to metallic electrodes were studied by several groups in
the absence of phonon thermal conductivity.[10-16] For the
applications of solid state coolers and power generators at room
temperature, one needs to consider a large number of serially
coupled QDs, otherwise it is not easy to maintain a large
temperature difference across the QD junction, which was pointed out
to be crucial in the implementation of high-efficiency
thermoelectric devices.[1,2] In addition, the phonon thermal
conductivity plays a significant role in the optimization of ZT at
high temperatures. In this paper, we carry out theoretical analysis
of $ZT$ of QD arrays embedded in nanowires, including the phonon
conductance, which is calculated by using the phonon radiative
transfer method as introduced in Ref. [17]. Although the method does
not take into account the microscopic mechanisms associated with
quantum confinement of acoustic phonons, it gives reasonable
agreement with the lattice-dynamics model and experiments for
nanowires by merely considering the boundary scattering effect of
phonons.

\section{Formalism}
%\textbf{2. Formalism}

A QDA embedded in a nanowire connected to the metallic electrodes
can be described by the Hurbbard-Anderson model. Here we consider
nanoscale semiconductor QDs, in which the energy level separations
are much larger than their on-site Coulomb interactions and thermal
energies. Thus, only one energy level for each quantum dot needs to
be considered. The Hamiltonian of the system is given by
$H=H_0+H_{QD}$:

\begin{eqnarray}
H_0& = &\sum_{k,\sigma} \epsilon_k
a^{\dagger}_{k,\sigma}a_{k,\sigma}+ \sum_{k,\sigma} \epsilon_k
b^{\dagger}_{k,\sigma}b_{k,\sigma}\\ \nonumber &+&\sum_{k,\sigma}
V_{k,L}d^{\dagger}_{L,\sigma}a_{k,\sigma}
+\sum_{k,\sigma}V_{k,R}d^{\dagger}_{R,\sigma}b_{k,\sigma}+c.c
\end{eqnarray}
where the first two terms describe the free electron gas of left and
right electrodes. $a^{\dagger}_{k,\sigma}$
($b^{\dagger}_{k,\sigma}$) creates  an electron of momentum $k$ and
spin $\sigma$ with energy $\epsilon_k$ in the left (right)
electrode. $V_{k,\ell}$ ($\ell=L,R$) describes the coupling between
the electrodes and the left (right) QD. $d^{\dagger}_{\ell,\sigma}$
($d_{\ell,\sigma}$) creates (destroys) an electron in the $\ell$-th
dot.

\begin{small}
\begin{eqnarray}
H_{QD}&=& \sum_{\ell,\sigma} E_{\ell} n_{\ell,\sigma}+
\sum_{\ell} U_{\ell} n_{\ell,\sigma} n_{\ell,\bar\sigma}\\
\nonumber &+&\frac{1}{2}\sum_{\ell,j,\sigma,\sigma'}
U_{\ell,j}n_{\ell,\sigma}n_{j,\sigma'}
+\sum_{\ell,j,\sigma}t_{\ell,j} d^{\dagger}_{\ell,\sigma}
d_{j,\sigma},
\end{eqnarray}
\end{small}
where { $E_{\ell}$} is the spin-independent QD energy level, and
$n_{\ell,\sigma}=d^{\dagger}_{\ell,\sigma}d_{\ell,\sigma}$.
$U_{\ell}$ and $U_{\ell,j}$ describe the intradot and interdot
Coulomb interactions, respectively. $t_{\ell,j}$ describes the
electron interdot hopping. Note that the interdot Coulomb
interactions as well as intradot Coulomb interactions play a
significant role on the charge transport for semiconductor QD array.

Using the Keldysh-Green's function technique [18], the charge and
heat currents of electrons leaving electrodes are expressed as

\begin{eqnarray}
J&=&\frac{2e}{h}\int d\epsilon {\cal T}(\epsilon)
[f_L(\epsilon)-f_R(\epsilon)],\\ Q_{L(R)}&=&\pm \frac{2}{h}\int
d\epsilon {\cal T}(\epsilon)(\epsilon-\mu_{L(R)})
[f_L(\epsilon)-f_R(\epsilon)],
\end{eqnarray}
where ${\cal T}(\epsilon)$ is the transmission coefficient.
$f_{L(R)}(\epsilon)=1/[e^{(\epsilon-\mu_{L(R)})/k_BT_{L(R)}}+1]$
denotes the Fermi distribution function for the left (right)
electrode. $\mu_L$ and $\mu_R$ denote the chemical potentials of the
left and right leads, respectively, with their average denoted by
$E_F=(\mu_L+\mu_R)/2$. $(\mu_L-\mu_R)=e\Delta V$ is the voltage
across the QDA junction. $T_{L(R)}$ denotes the equilibrium
temperature of the left (right) electrode. $e$ and $h$ denote the
electron charge and Planck's constant, respectively. $Q_{L(R)}$
denotes the heat current leaving from the left (right) electrode.

In the linear response regime, Eqs.~(3) and (4) can be rewritten as
\begin{eqnarray}
J&=&{\cal L}_{11} \frac{\Delta V}{T}+{\cal L}_{12} \frac{\Delta
T}{T^2}\\Q&=&{\cal L}_{21} \frac{\Delta V}{T}+{\cal L}_{22}
\frac{\Delta T}{T^2},
\end{eqnarray}
where there are two sources of driving force to yield the charge and
heat currents. $\Delta T=T_L-T_R$ is the temperature difference
across the junction.  The thermoelectric coefficients in Eqs.~(5)
and (6) (${\cal L}_{11}$, ${\cal L}_{12}$, ${\cal L}_{21}$, and
${\cal L}_{22}$) are evaluated by

\begin{equation}
{\cal L}_{11}=\frac{2e^2T}{h} \int d\epsilon {\cal T}(\epsilon)
(\frac{\partial f(\epsilon)}{\partial E_F})_T,
\end{equation}
\begin{equation}
{\cal L}_{12}=\frac{2eT^2}{h} \int d\epsilon {\cal T}(\epsilon)
(\frac{\partial f(\epsilon)}{\partial T})_{E_F},
\end{equation}

\begin{equation}
{\cal L}_{21}=\frac{2eT}{h} \int d\epsilon {\cal
T}(\epsilon)(\epsilon-E_F) (\frac{\partial f(\epsilon)}{\partial
E_F})_T,
\end{equation}
and
\begin{equation}
{\cal L}_{22}=\frac{2T^2}{h} \int d\epsilon {\cal T}(\epsilon)
(\epsilon-E_F)(\frac{\partial f(\epsilon)}{\partial T})_{E_F}.
\end{equation}
Here ${\cal T}(\epsilon)$ and
$f(\epsilon)=1/[e^{(\epsilon-E_F)/k_BT}+1]$ are evaluated under the
equilibrium condition. The detailed expression of ${\cal
T}(\epsilon)$ can be found in Ref. [19].

If the system is in an open circuit, the electrochemical potential
will be established in response to a temperature gradient; this
electrochemical potential is known as the Seebeck voltage (Seebeck
effect). The Seebeck coefficient (amount of voltage generated per
unit temperature gradient) is defined as $S=\Delta V/\Delta T=-{\cal
L }_{12}/(T{\cal L}_{11})$. To judge whether the system is able to
generate or extract heat efficiently, we need to consider the figure
of merit [1]
\begin{eqnarray}
ZT=\frac{S^2G_eT}{\kappa_e+\kappa_{La}}\equiv
\frac{(ZT)_0}{1+\kappa_{La}/\kappa_e}.
\end{eqnarray}
Here, $G_e={\cal L}_{11}/T$ is the electrical conductance and
$\kappa_e= (({\cal L}_{22}/T^2)-S^2 G_e T)$ is the electron thermal
conductance. $(ZT)_0$ represents the $ZT$ value in the absence of
phonon thermal conductance, $\kappa_{La}$. In the Hamiltonian $H$,
the term ($H_{e,ph}$) describing the interactions between electrons
and phonons is ignored. For simplicity, we adopt
$\kappa_{La}=\kappa_{wire} F_s$ to describe the phonon thermal
conductance of QDA embedded a nanowire. The dimensionless scattering
factor $F_s$ is used to include the phonon scattering effect arising
from surface boundary of QDs.[1] It is possible to reduce phonon
thermal conductance by one order of magnitude when QDs in a quantum
wire behave like phonon scatterers.[20] Therefore, the maximum $F_s$
is assumed to be 0.1 in this study. For example, Ref. [20] pointed
out that the phonon thermal conductance reduction of QDA nanowires
in a high temperature regime arises from the filtering of high
frequency phonons.[2] However, the method considered in Ref. [20]
requires a heavily numerical calculation of $\kappa_{La}$ in the
optimization of $ZT$.

Based on Holland's model [21,22] in which the longitudinal and
transverse acoustical phonon branches are treated separately, we
calculate $\kappa_{wire}$ of a rectangular nanowire with  cross
sectional area $A$ and length $L_x$. The thermal conductance
$\kappa_{wire}$ is related to the phonon thermal conductivity by
$\kappa_{wire}=\kappa_{ph}A/L_x$, where
$\kappa_{ph}=K_{T_0}+K_{Tu}+K_{L}$ denotes the thermal conductivity
with
\begin{eqnarray}
K_{T_0}=\frac{2}{3}T^3\int^{\theta_{1}/T}_{0}\frac{C_{T_0}x^4e^x(e^x-1)^{-2}dx}{\tau_b^{-1}+\tau^{-1}_I+\beta_{T}xT^5},
\end{eqnarray}

\begin{eqnarray}
K_{Tu}=\frac{2}{3}T^3\int^{\theta_{2}/T}_{\theta_{1}/T}\frac{C_{Tu}x^4e^x(e^x-1)^{-2}dx}{\tau_b^{-1}+\tau^{-1}_I+\beta_{Tu}x^2T^2/\sinh(x)},
\end{eqnarray}
and
\begin{eqnarray}
K_{L}=\frac{1}{3}T^3\int^{\theta_{3}/T}_{0}\frac{C_{L}x^4e^x(e^x-1)^{-2}dx}{\tau_b^{-1}+\tau^{-1}_I+\beta_{L}x^2T^5}.
\end{eqnarray}
The phonon thermal conductivities  $K_{T_0}$, $K_{Tu}$, and $K_{L}$
result from the low-frequency transverse branch, high-frequency
transverse branch, and longitudinal acoustical phonon branch,
respectively. In Eqs.~(12)-(14), we have $C_{j=T_0,Tu,
L}=(k_B/2\pi^2v_j)(k_B/\hbar)^3$, where $v_j$ is the phonon group
velocity for the $j^{th}$-branch. $\tau_b^{-1}$ is the phonon
boundary scattering rate, and $\tau^{-1}_I=\alpha x^4T^4$ is the
phonon-impurity scattering rate. $\beta_TxT^5$,
$\beta_{Tu}x^2T^2/\sinh(x)$, and $\beta_{L}x^2T^5$ arise from
three-phonon scattering rates. $\alpha$ and $\beta_j$ are empirical
parameters determined by fitting experimental data.
$\theta_i(\hbar\omega_i/k_B)(i=1,2,$ and $3)$ denotes the Debye
temperature. The frequencies of $\omega_i$, for $T_0$, $Tu$ and $L$
modes can be found in Ref. [17]. The $\alpha$, $\beta_j$, $v_j$ and
$\theta_i$ parameters for Si and GaAs are adopted from Refs. [17]
and [22], respectively. Eqs. (12)-(14) can well describe the phonon
thermal conductivity of semiconductor materials with $\tau_b=v_b/L$.
$v_b(=[(2v_T^{-1}+v_L^{-1})/3]^{-1})$ is the average phonon group
velocity, where $v_T$ and $v_L$ correspond to the group velocities
of the transverse and longitudinal branches. $L$ denotes the sample
size.  Chen and Tien [21] extended the Holland model to illustrate
the phonon thermal conductivity of quantum wells by considering the
geometry effect on $\tau_b$. Based on the formalism of ref.[21], the
phonon-boundary scattering rate ($\tau_b^{-1}$) of quantum wire is
derived and determined by

\begin{eqnarray}
\tau_b^{-1}=(1/{\cal G}-1)\cdot \tau_t^{-1},
\end{eqnarray}
where $\tau_t=(A\omega^4+(B_1+B_2)T^3\omega^2+v_b/L_c)^{-1}$ is the
total internal relaxation time of the bulk which includes normal
process and umklapp process. $A=\alpha (\hbar/k_B)^4$, and
$B_j=\beta_j (\hbar/k_B)^2$. $L_c$ denotes the sample length. The
factor of ${\cal G}={\cal G}^{++}+{\cal G}^{+-}+{\cal G}^{-+}+{\cal
G}^{--}$ illustrates the boundary effects on the phonon thermal
conductivity of nanowires. The expression of ${\cal G}$ is given by

(i) $0<\theta<\pi/2$ and $0<\phi<\pi$
\begin{eqnarray}
{\cal G}^{++}/\Pi=\int d\Omega
\frac{(e^{-\xi_y}-1)(e^{-\xi_z}-1)}{e^{-\xi_y}e^{-\xi_z}-1},
\end{eqnarray}

(ii) $0<\theta<\pi/2$ and $\pi<\phi<2\pi$
\begin{eqnarray}
{\cal G}^{+-}/\Pi=\int d\Omega
\frac{(e^{-\xi_y}-1)(e^{-\xi_z}-1)}{e^{-\xi_y}e^{-\xi_z}-1},
\end{eqnarray}

(iii) $\pi/2<\theta<\pi$ and $0<\phi<\pi$
\begin{eqnarray}
{\cal G}^{-+}/\Pi=\int d\Omega
\frac{(e^{-\xi_y}-1)(e^{-\xi_z}-1)}{e^{-\xi_y}e^{-\xi_z}-1},
\end{eqnarray}

(iv) $\pi/2<\theta<\pi$ and $\pi<\phi<2\pi$
\begin{eqnarray}
{\cal G}^{--}/\Pi=\int d\Omega
\frac{(e^{-\xi_y}-1)(e^{-\xi_z}-1)}{e^{-\xi_y}e^{-\xi_z}-1}.
\end{eqnarray}
We have the notations $\Pi=\frac{3}{4\pi L_y L_z}$ and $d\Omega=
dydzd\theta d\phi~\sin^3(\theta)~cos^2(\phi)$ in Eqs.~(16)-(19),
where $L_{y(z)}$ is the lateral size of rectangular wire in the y(z)
direction. The other notations are $\xi_y=-y/(\Lambda \sin(\theta)
\sin(\phi))$, $\xi_z=-z/(\Lambda \cos(\theta))$,
$\xi_y=(L_y-y)/(\Lambda \sin(\theta) \sin(\phi))$ and
$\xi_z=(L_z-z)/(\Lambda \cos(\theta))$. The average phonon mean free
path is assumed to be $\Lambda=v_b\tau_t$.

The phonon thermal conductivities ($\kappa_{ph}$) of rectangular Si
and GaAs nanowires calculated by using Eqs.~(12)-(14) are plotted
for two different topological structures in Fig. 1. The calculations
of silicon nanowire can be used to compare with experimental results
to examine the validity of the scheme adopted in Ref. [21]. The heat
problem of high efficiency solar cells made of III-V compounds (such
as GaAs) can degrade the system performance.[1] Thus, it is
important to design a GaAs cooler, which can be integrated with the
solar cell to solve such a problem. Figure 1(a) shows $\kappa_{ph}$
of Si nanowire with $L_y =L_z$ as a function of temperature for
three cross-section areas: solid line ($A=(30~nm)^2$), dashed-dotted
line ($A=(10~nm)^2$) and dashed line ($A=(5~nm)^2$). It is expected
that $\kappa_{ph}$ decreases with decreasing cross-section area due
to enhanced boundary scattering. For a nanowire with small cross
section, $\kappa_{ph}$ increases slower with respect to $k_BT$ for
$T> 200~K$. These results are consistent with other theoretical
calculations [23] and experimental observations.[24,25] For
nanowires with large cross section, however, $\kappa_{ph}$  is
slightly underestimated compared to other theoretical works.[23] For
the smallest area considered ($A=(5~nm)^2$), the $\kappa_{ph}$ value
obtained is very close to the result calculated from the
lattice-dynamics model in Ref. 20, which takes into account the
quantum confinement of acoustic phonons. Although a nanowire with
even smaller cross section leads to much smaller phonon thermal
conductivity, it is a challenge to implement such tiny
nanowires.[1,2]

Figure~1(b) shows $\kappa_{ph}$ of GaAs nanowire as a function of
temperature for different $L_z$ sizes with  $L_y$ fixed at $5~nm$.
We note that $\kappa_{ph}$ of GaAs is smaller than that of Si in a
wide temperature range for $A=(5~nm)^2$ because the average group
velocity in GaAs is smaller. $\kappa_{ph}$ increases with increasing
$L_z$. However, $\kappa_{ph}$ for $L_z=500~nm$ becomes almost the
same as $\kappa_{ph}$ of GaAs thin film with $L_z=L_c=0.729 cm$ and
$L_y=5~nm$ (see the solid and black line). This implies that the
size effect of $L_z$ on $\kappa_{ph}$ can be ignored when $L_z$ is
larger than $500~nm$, {\color{red}comparable to} the average phonon
mean free path of $GaAs$. The maximum $\kappa_{ph}$ value of GaAs
thin film at $T=100~K$ is around $4~W/mK$, which is still smaller
than that of  Si nanowire with $A=(10~nm)^2$. This implies that
$GaAs$ may have a better $ZT$ value than silicon at room
temperature. Our results for $GaAs$ are consistent with the
calculation of Ref.~[26]. Although the TE properties of
germanium/silicon QDs system are interesting for their low cost
fabrication process [27], the silicon semiconductor TE devices can
not be directly integrated with III-V compound solar cell
systems.[1] Therefore, we focus on the $ZT$ optimization of
InGaAs/GaAs QDs junction system in the next section.

\section{Results and discussion}
%\textbf{3. Results and discussion}

%\subsection{Triple quantum dots}
Because it is important to reduce the phonon thermal conductivity in
the optimization of $ZT$, we need to consider a QDA nanowire with
long length. Therefore, before investigating the TE properties of
QDAs with large number of coupled dots, we first consider the
simplest structure of QDA, which consists of serially coupled triple
QDs (SCTQD). Although we have theoretically studied the transport
and TE properties of triple QD molecules with full solution in the
Coulomb blockade regime,[28,29] it is still a challenging work to
obtain the full solution to the charge transport through QDAs with
more than three dots due to the fact that the calculation scales up
exponentially. To solve such a difficult problem many efforts which
adopt some approximation schemes were attempted.[9,30-32] The
Hartree-Fock approximation (HFA) is often used to investigate the TE
coefficients of realistic molecules.[30-32] As a consequence, it is
desirable to examine the validity of such a mean-field approach. At
high temperatures, in the absence of interdot Coulomb interactions
the results based on the approach of Ref. [9] can achieve good
agreement with those calculated by the full solution as reported in
Ref. [29]. Therefore, we can examine the validity of HFA method by
using the approach of Ref.[9] instead of using the full solution of
Ref. [29], which requires much heavier numerical calculation. In
addition, the approximation of Ref. [9] can provide a closed form
transmission coefficient of Eqs. (7)-(10), which is useful for
analyzing the charge transport properties.

Because the optimization of power factor (PF) favors identical QDs,
size-independent electron hopping strength, and symmetrical
tunneling rates,[9] in Fig.~2 we plot $G_e$, $S$, $\kappa_e$ and
$PF$ of SCTQD as functions of gate voltage ($eV_g$) at two different
temperatures by considering $t_{\ell,j}=t_c=3\Gamma_0$,
$U_{\ell}=U_0=30\Gamma_0$ and symmetrical tunneling rates
($\Gamma_L=\Gamma_R=1\Gamma_0$). All energy scales are in units of
$\Gamma_0$, which is taken to be $1~meV$. Here, the interdot Coulomb
interactions are ignored for simplicity when we examine the validity
of HFA method. The solid lines and dot-dashed lines are calculated
according to the procedure of Ref.[9] and the HFA method of
Ref.[30-32], respectively. The approach of Ref.~[9], which neglects
the interdot Coulomb interactions, is called the on-site Coulomb
approximation (OCA). Both OCA and HFA are efficient tools for
studying the charge transport through molecular junctions.

For all {four physical quantities ($G_e$, $S$, $\kappa_e$ and
$PF=S^2G_e$),} the spectra consist two main groups of features,
which are mirror image of each other with respect to the mid point
($eV_g=45\Gamma_0$). This is a consequence of the electron-hole
symmetry possessed by the SCTQD structures in the absence of
interdot Coulomb interactions [29]. At low temperature
($k_BT=0.1\Gamma_0$), there is a gap between the two groups, which
is related to on-site Coulomb interaction, $U_0$. The first three
peaks of $G_e$ (shown by the solid line) correspond to three
resonant channels at $E_0-\sqrt{2}t_c$, $E_0$, and
$E_0+\sqrt{2}t_c$. For the higher-energy group, there are also three
peaks arising from resonant channels at $E_0+U_0-\sqrt{2}t_c$,
$E_0+U_0$ and $E_c+U_0+\sqrt{2}t_c$. We find that the behaviors of
$G_e$, $S$ and $\kappa_e$ calculated by HFA (dot-dashed lines) are
quite different from those calculated by OCA (solid lines). For
example, the maximum $G_e$ of dot-dashed lines is larger than
$e^2/h$ (quantum conductance), but not for OCA. This is due to the
inability of HFA to include electron correlation. The energy
positions of resonant channels within HFA depend on the average
occupation number in each QD.[30-32], which can be a fractional
number. In  contrast, within OCA  the energy positions of resonant
channels are determined by integer charges, while electron
correlation functions and fractional occupation numbers only appear
in the probability weights of quantum paths.[9] In addition, we
observe an unphysical enhancement of $S$ in the middle of the
Coulomb gap $(eV_g=45\Gamma_0)$ for HFA results (dot-dashed line).
This illustrates the drawback of HFA calculation for TE coefficients
of QDs at low temperatures when the charging process plays a
significant role. At high temperatures (for example,
$k_BT=10\Gamma_0$), the detailed resonant structures are washed out,
and the unphysical enhancement of $S$ in the middle of the Coulomb
gap also disappears. These two approaches give qualitatively similar
results at high temperatures. However, the HFA approach tends to
overestimate $G_e$,  $\kappa_e$, and  $PF$. If we include the
interdot Coulomb interactions, the interdot correlation effects
become crucial[29]. {\color{red}The symmetrical behavior of TE
coefficients will be lost. In general, we will obtain smaller PF for
QD energy levels below $E_F$ than that for QD energy levels above
$E_F$.[9,29] This indicates that to achieve maximum PF, it is
preferable to have the orbital depletion situation with the total
occupation number $N=\sum_{\sigma}
(N_{L,\sigma}+N_{C,\sigma}+N_{R,\sigma})\le 1$, where
$N_{\ell,\sigma}$ denotes the single particle occupation number in
dot $\ell$.[29]}

For TE devices operated at room temperature, it is important to
optimize their $ZT$ values at high temperatures. To further clarify
the differences between these two approaches, Fig. 3 shows
$G_e$, $S$, $\kappa_e$ and $PF$ as functions of the detuning energy
($\Delta=E_0-E_F > 0 $) at various temperatures. {\color{red}Note
that in Fig. 3, we consider the case with QD energy levels above $E_F$, which
satisfies the condition for orbital depletion. In one-particle
transport process (QD molecule with an empty state) only the
intradot Coulomb interaction effects are important.[29] This is a
typical feature for carrier transport in the Coulomb blockade
regime.[9]} The other physical parameters are the same as those of
Fig.~2. From the results of $S$ and $PF$, these two approaches agree
well with each other when QD energy levels are far above $E_F$. This
is expected, because the electron Coulomb interactions become
unimportant for large $\Delta$ values. The maximum PF value
decreases with increasing temperature. For $k_BT=1\Gamma_0$,
$k_BT=5\Gamma_0$ and $k_BT=13\Gamma_0$, the maximum $PF$ occurs
at near $\Delta=5\Gamma_0$, $\Delta=15\Gamma_0$, and
$\Delta=35\Gamma_0$, respectively. When $k_BT=26\Gamma_0$, the maximum PF occurs
at near $\Delta=50\Gamma_0$.

%\subsection{A single coupled QD chain with N=10}
From results of Fig.~3, we see that HFA is a reasonably good
approximation for analyzing the TE coefficients of QDAs with QD
energy levels above $E_F$. Thus, we can use HFA to investigate TE
coefficients of QDA nanowire with large numbmer of dots. Here, we
use  HFA (instead of OCA) of Ref.[30] to study QDA nanowire with
N=10 because of its simplicity.  Fig.~4 shows the $G_e$, $S$ and
$(ZT)_0$ as functions of temperature for various detuning energies
($\Delta$). The other physical parameters are the same as those of
Fig.~2. The behavior of $G_e$ with respective to temperature can be
understood by considering $G_e \approx 1/(k_BT~cosh^2(\Delta/2k_BT)$
for $U_{\ell}=0$ (see the curve with triangle marks). $S$ is
proportional to $T$ when temperature approaches zero (not shown
here), whereas $S\approx \Delta/k_BT$ at high temperatures and
$\Delta/\Gamma \gg 1$ (see the curves with triangle marks). In the
absence of $\kappa_{La}$, a very large $(ZT)_0$ appears in the low
temperature regime. The maximum $(ZT)_0$ is beyond three, which is a
critical value for the realization of solid state coolers and power
generators. $(ZT)_0$ larger than 3 is a direct consequence of the
{\color{red}fact $G_e\gg\kappa_e$ ($\kappa_e$ not shown here) and large
enhancement of $S^2$ in the Coulomb blockade regime. Using small
nanowires in the quantum thermal conductance regime and a vacuum space to
blockade surrounding  phonon heat current was proposed to yield a vanishingly} small
phonon thermal conductivity.[33] However, such a design requires
advanced nanotechnology. The results of Fig.~4(c) also imply that
the on-site electron Coulomb interactions can be ignored only when
$U_{\ell}/k_BT \gg 1$. $(ZT)_0$ is highly overestimated at high
temperatures in the absence of $U_{\ell}$.

Fig.~5(a) shows $ZT$ in the presence of
$\kappa_{La}$ {\color{red}($\kappa_{La}=F_s\kappa_{ph}(A/L_x)$)} for the case of a GaAs nanowire
with $L_z=L_y=5~nm$ and $L_x=250~nm$. $F_s$ is assumed to be 0.1 in
the calculation. The curves of Fig.~5(a) have one-to-one
correspondence to those of Fig.~4(c). The maximum $ZT$ value is
significantly reduced in the presence of $\kappa_{La}$. This reduction
of $ZT$ results from the fact $\kappa_{La}\gg\kappa_e$ [see Eq.~(11)],
which implies that the optimization of $ZT$ can be achieved by
reducing $\kappa_{ph}$ and increasing power factor $PF=S^2 G_e$
simultaneously. The maximum $ZT$ depends on the detuning energy,
$\Delta$. This behavior can be understood from the results of
Fig.~3(d). The $ZT$ value is almost independent of temperature for
$\Delta=40\Gamma_0$, when $k_BT \ge 20 \Gamma_0$. This is because
the ratio of $G_e/(T\kappa_{La})$ is nearly constant. In the high
temperature regime we have $\kappa_{ph}\approx 1/T$ and $G_e$ decays
very slowly with increasing temperature.

Fig.~5(b) shows $ZT$ as a function of temperature for various
electron hopping strengths at $\Delta=40\Gamma_0$. $ZT$ is
enhanced with increasing $t_c$. Under the condition
$\kappa_{La}\gg\kappa_e$, the behavior of $ZT$ with respect to
$t_c$ can be analyzed by the power factor $PF=S^2 G_e$. When the QD
energy is far above $E_F$ ($\Delta=40\Gamma_0$), the behavior of
$S=\Delta/k_BT$ is independent of $t_c$ in the Coulomb blockade
regime once $k_BT \ge 10\Gamma_0$. Therefore, the enhancement of
$ZT$ is completely determined by the increase of $G_e$ with respect
to $t_c$. When $t_c$ is smaller than the tunneling rate
($\Gamma_L=\Gamma_R=\Gamma$), $G_e$ increases quickly with
increasing $t_c$, and it becomes insensitive to $t_c$ when $t_c \ge
2\Gamma_0.$ This demonstrates that the optimization of $ZT$ can be
realized by the enhancement of PF for a given $\kappa_{La}$.

In Fig.~5, we consider square nanowires with a small cross section
with $L_y=L_z=5~nm$. To understand the dependence on QDA
cross-sectional area, we show in Fig.~6 $ZT$ as a function of
temperature for various values  of $L_z$ with $L_y$ fixed at $5~nm$.
We adopt $\Gamma_L=\Gamma_R=\Gamma=3\Gamma_0$, $t_c=3.5\Gamma_0$,
$U_{\ell}=30\Gamma_0$ and $\Delta=40\Gamma_0$. In our model, we keep
a large separation between QDAs along the z direction ($D_s=25~nm$),
this allow us to avoid the electron hopping effects and
{\color{red}interdot} Coulomb interactions between QDAs along that
direction. Fig.~6 shows the maximum $ZT$ of QDA with $L_z=7L_y$ can
be larger than one. This is due to the enhancement of $G_e$ with
increased tunneling rates ($\Gamma=3\Gamma_0$). The $ZT$ value
reduces significantly for the cases with large  $L_z$ ($\ge 13L_y$).
This $ZT$ reduction comes from the increased value of $\kappa_{La}$
[see Fig.~1(b)]. If we reduce the separation, $D_s$ between QDAs to
increase the density of QDAs, the proximity effect arising from
{\color{red}interdot} Coulomb interactions between QDAs should be
included.[33] In general, the proximity effect from interdot Coulomb
interactions will suppress the $G_e$ values.

\section{Summary and conclusions}
%\textbf{4. Summary and conclusions }

We have theoretically investigated the TE properties of QDAs
embedded in nanowires connected to metallic electrodes in the
Coulomb blockade regime. Both OCA and HFA calculations were
examined. It was shown that HFA method gives unphysical results for
the Seebeck coefficient at low temperature when the QDs are nearly
half filled. However, it gives reasonable results at high
temperatures when the QD energy levels are far above the Fermi level
(low filling condition). Under the condition $\kappa_{La} \gg
\kappa_e$, the optimization of $ZT$ can be achieved by reducing
$\kappa_{La}$ and increasing the power factor $PF=S^2G_e$,
simultaneously. We found that for various design parameters, $ZT>1$
can be achieved. In our simulation, we assumed that the presence of
QDs reduces the phonon conductivity in nanowires by a factor
$F_s=0.1$. It is conceivable that a smart nanostructure design which
reduces the value of $F_s$ further can increase the maximum $ZT$
even more. The maximum $ZT$ value of a QDA nanowire with small cross
section is found insensitive to temperature variation (see
Figs.~5(b) and 6). Such a feature is quite different from that of
conventional thermoelectric materials.[1,2] The ability of QDAs to
maintain $ZT$ near its maximum value for a wide range of
temperatures is an important feature in the realization of useful
thermoelectric devices.[2]

\begin{flushleft}
{\bf Acknowledgments}\\
\end{flushleft}
This work was supported in part by the National Science Council of
the Republic of China under Contract Nos. NSC 103-2112-M-008-009-MY3
and NSC 101-2112-M-001-024-YM3.

\mbox{}\\
${}^{\dagger}$ E-mail address: mtkuo@ee.ncu.edu.tw\\
${}^{*}$ E-mail:yiachang@gate.sinica.edu.tw

%\mbox{}

%\newpage

%\textbf{Figure caption }

\clearpage
%\begin{figure}[h]
%\centering
%\includegraphics[angle=-90,scale=0.5]{Fig1}
%\caption{(a)One dimensional quantum dot array connected to
%electrodes, and (b) two dimensional quantum dot array connected to
%electrodes.}
%\end{figure}

\begin{figure}[h]
\centering
\includegraphics[angle=-90, scale=0.3]{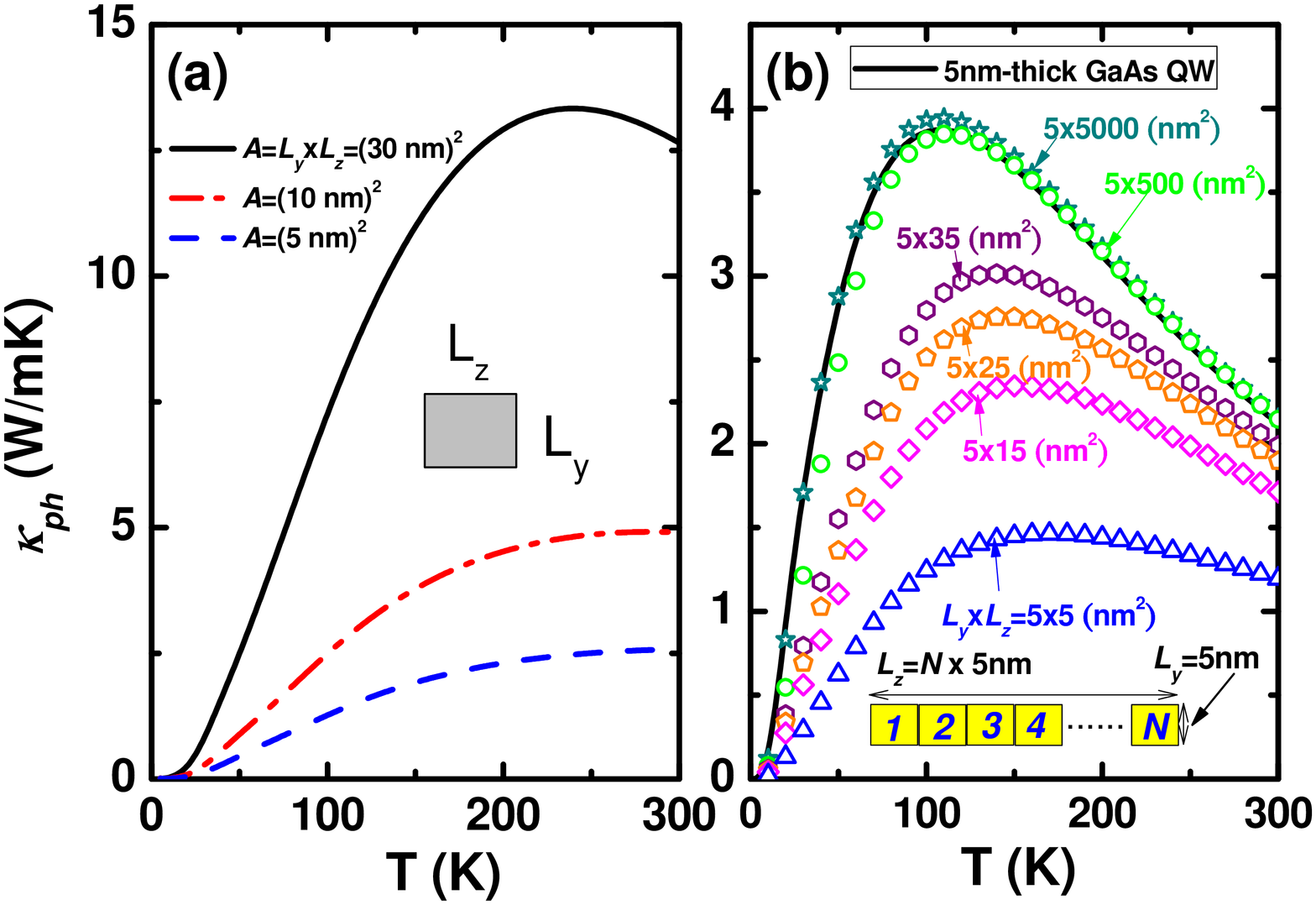}
\caption{(a) Phonon thermal conductivity of a silicon nanowire as a
function of temperature for the different cross sections. The square
cross section is defined as $A=L_y \times L_z$, where $L_y$ and
$L_z$ are lateral sizes, and (b) phonon thermal conductivity of a
GaAs nanowire with rectangular cross section $A=L_y\times L_z$ as a
function of temperature for different $L_z$ values at fixed
$L_y=5~nm$.}
\end{figure}

\begin{figure}[h]
\centering
\includegraphics[angle=-90,scale=0.3]{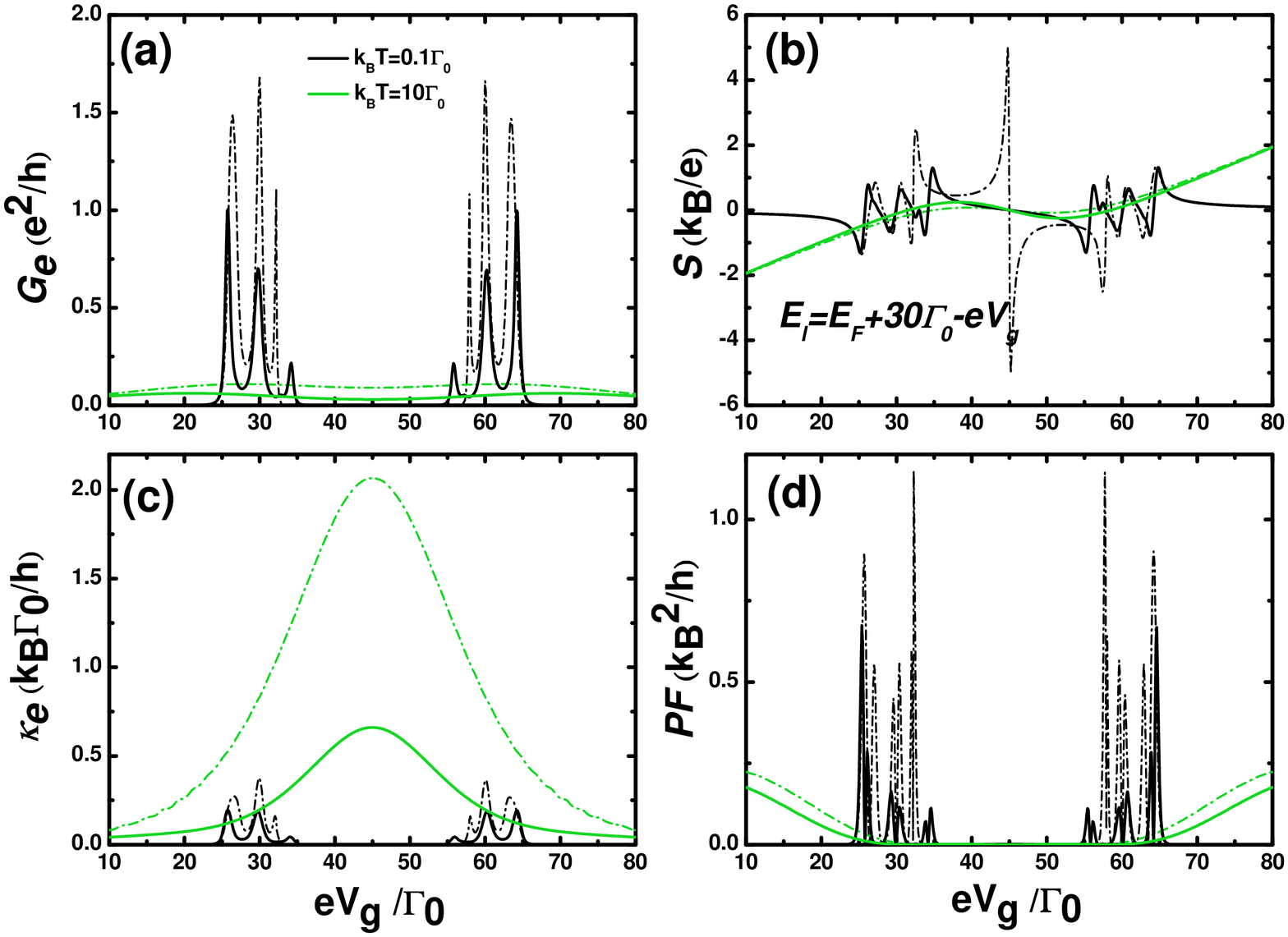}
\caption{(a) Electrical conductance ($G_e$), (b) Seebeck coefficient
(S), (c) electron thermal conductance ($\kappa_e$), and (d) power
factor $PF=S^2G_e$ of triple QDs junction systems as a function of
gate voltage for two different temperatures. Solid line and
dot-dashed line are calculated by  OCA and
HFA, respectively.}
\end{figure}

\begin{figure}[h]
\centering
\includegraphics[angle=-90,scale=0.3]{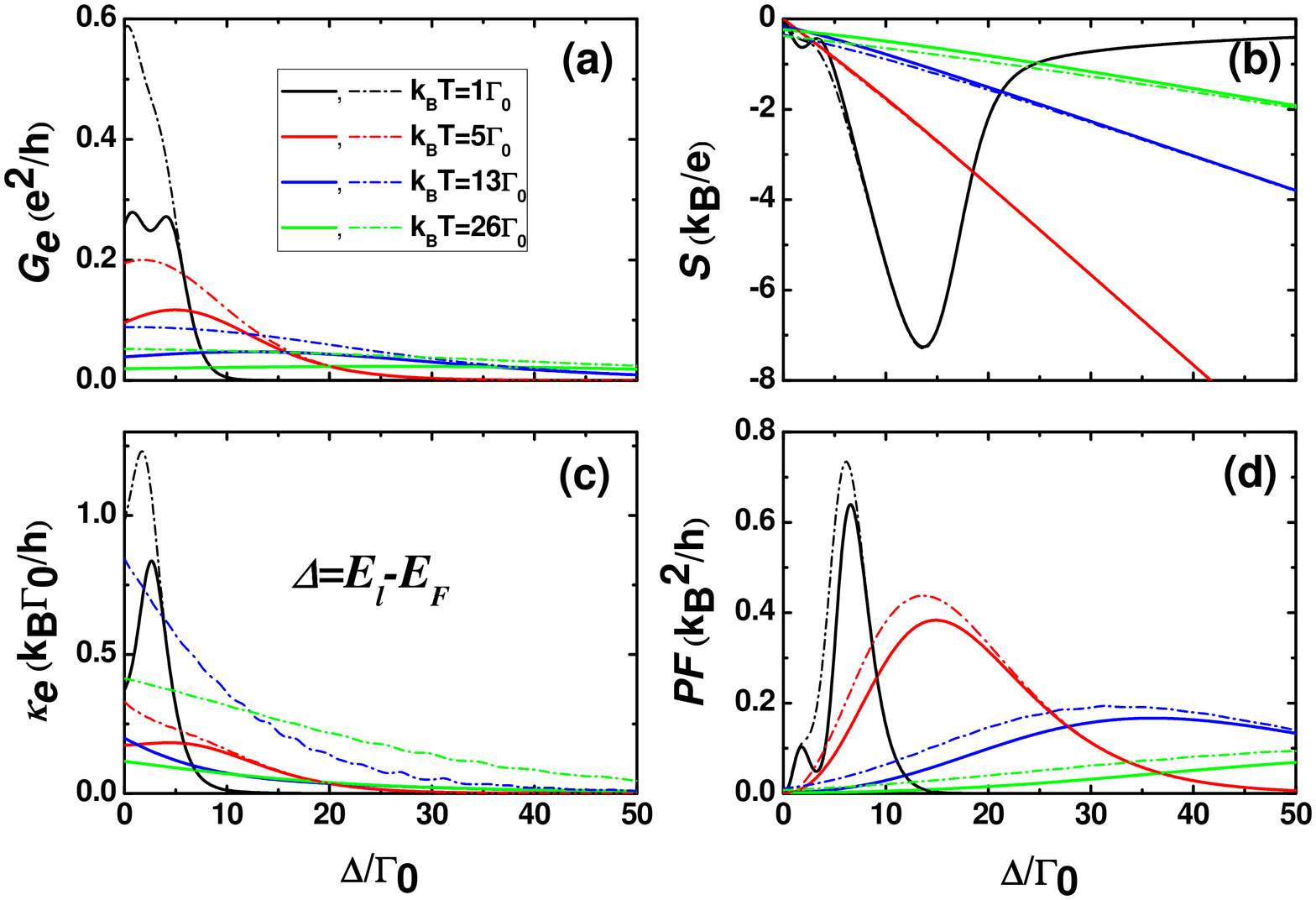}
\caption{(a) Electrical conductance ($G_e$), (b) Seebeck coefficient
(S), (c) electron thermal conductance ($\kappa_e$), and (d) power
factor $PF=S^2G_e$ of triple QDs junction systems as a function of
detuning energy $\Delta=E_{0}-E_F$ for different temperatures. The
other physical parameters are the same as those of Fig. 2.}
\end{figure}

\begin{figure}[h]
\centering
\includegraphics[angle=-90,scale=0.3]{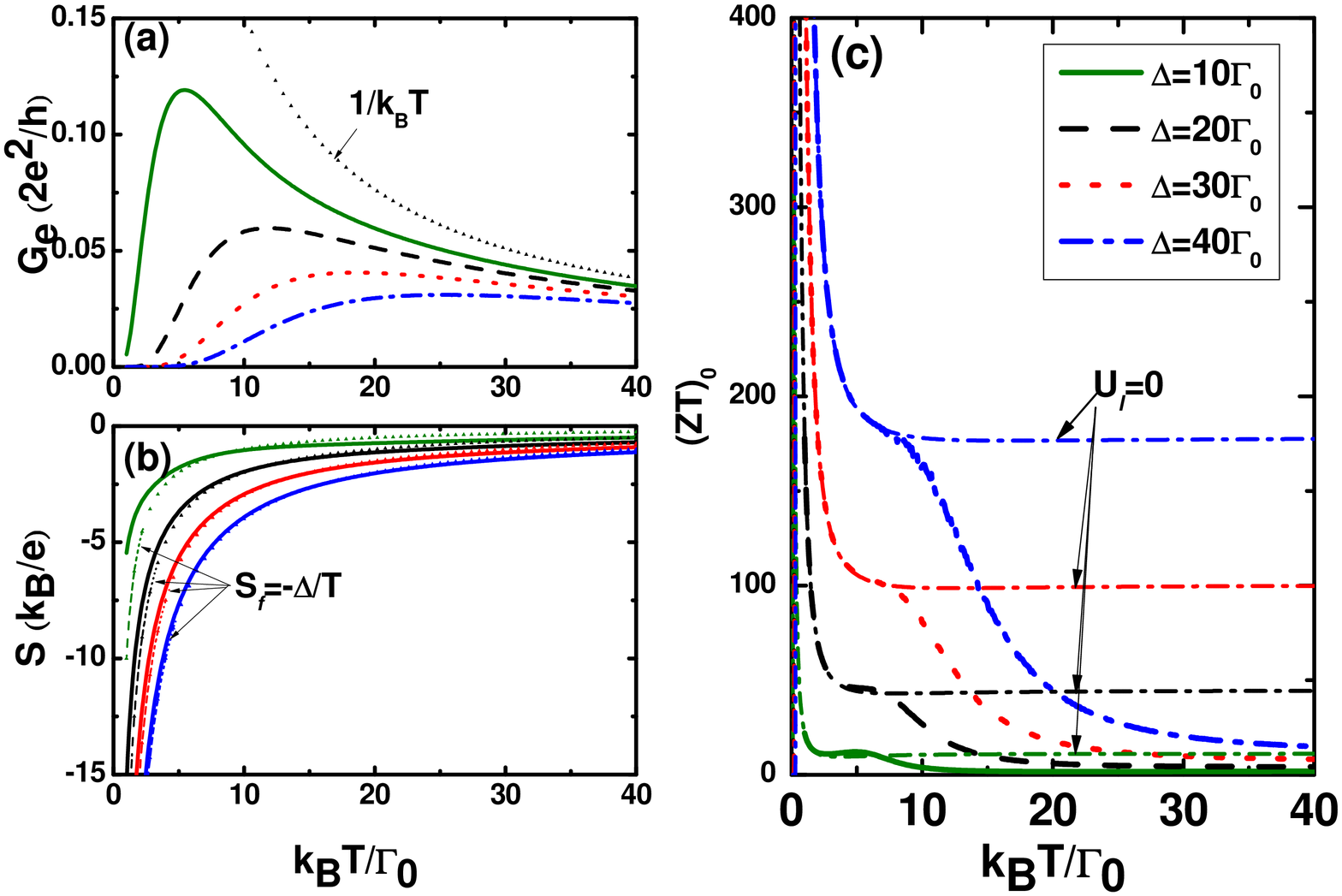}
\caption{(a) Electrical conductance, (b) Seebeck coefficient and (c)
$(ZT)_0$ of a QDA system with dot number N=10 in the absence of
$\kappa_{ph}$ as a function of temperature for different detuned
energies.  Other physical parameters are the same as those of Fig.2.}
\end{figure}

\begin{figure}[h]
\centering
\includegraphics[angle=-90,scale=0.3]{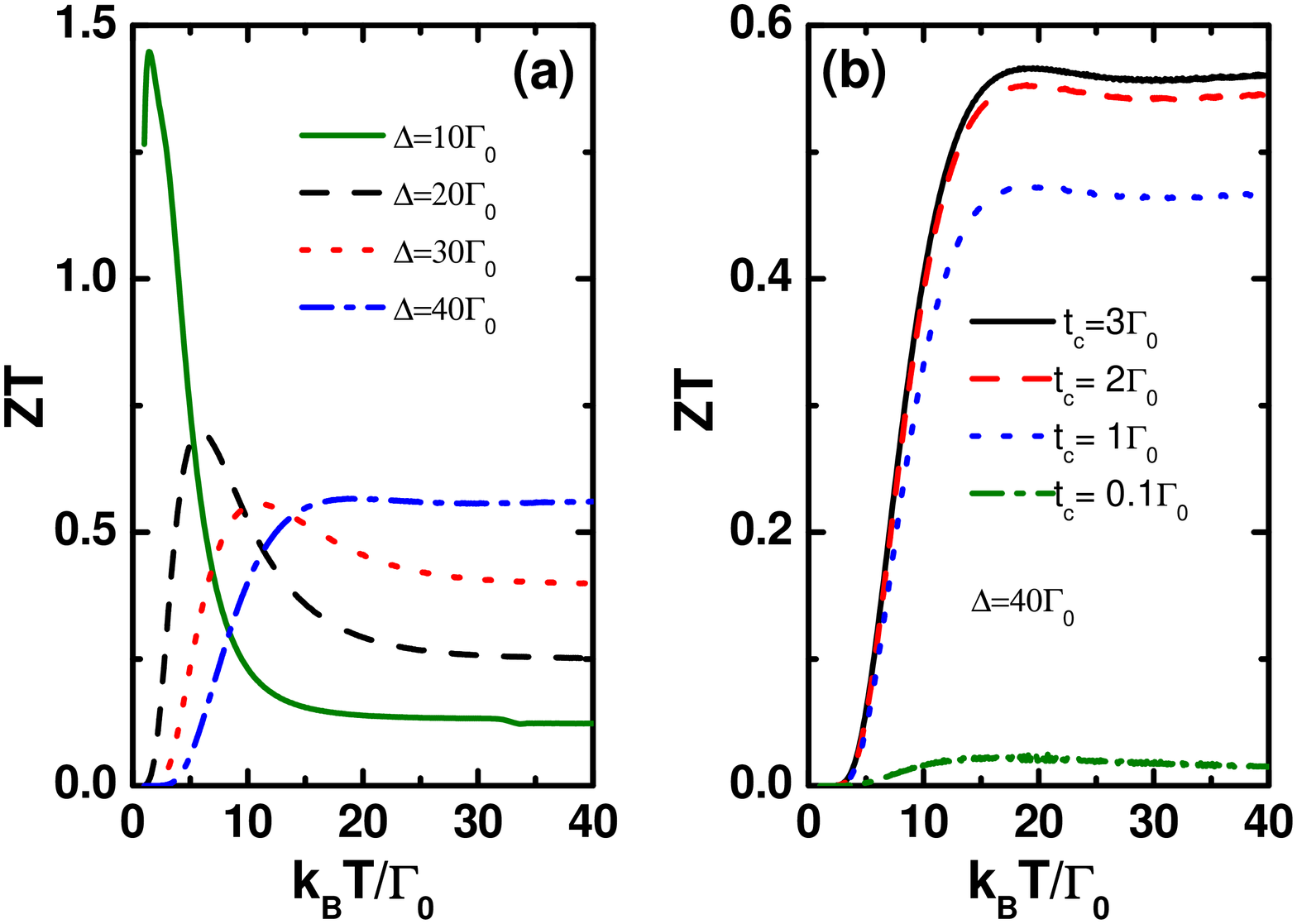}
\caption{Figure of merit of a QDA system with dot number N=10 in the
presence of $\kappa_{ph}$ as a function of temperature. The curves
of diagram (a) correspond to those of Fig. 4. Diagram (b) shows
$ZT$ for various electron hopping strengths. }
\end{figure}

\begin{figure}[h]
\centering
\includegraphics[angle=-90,scale=0.3]{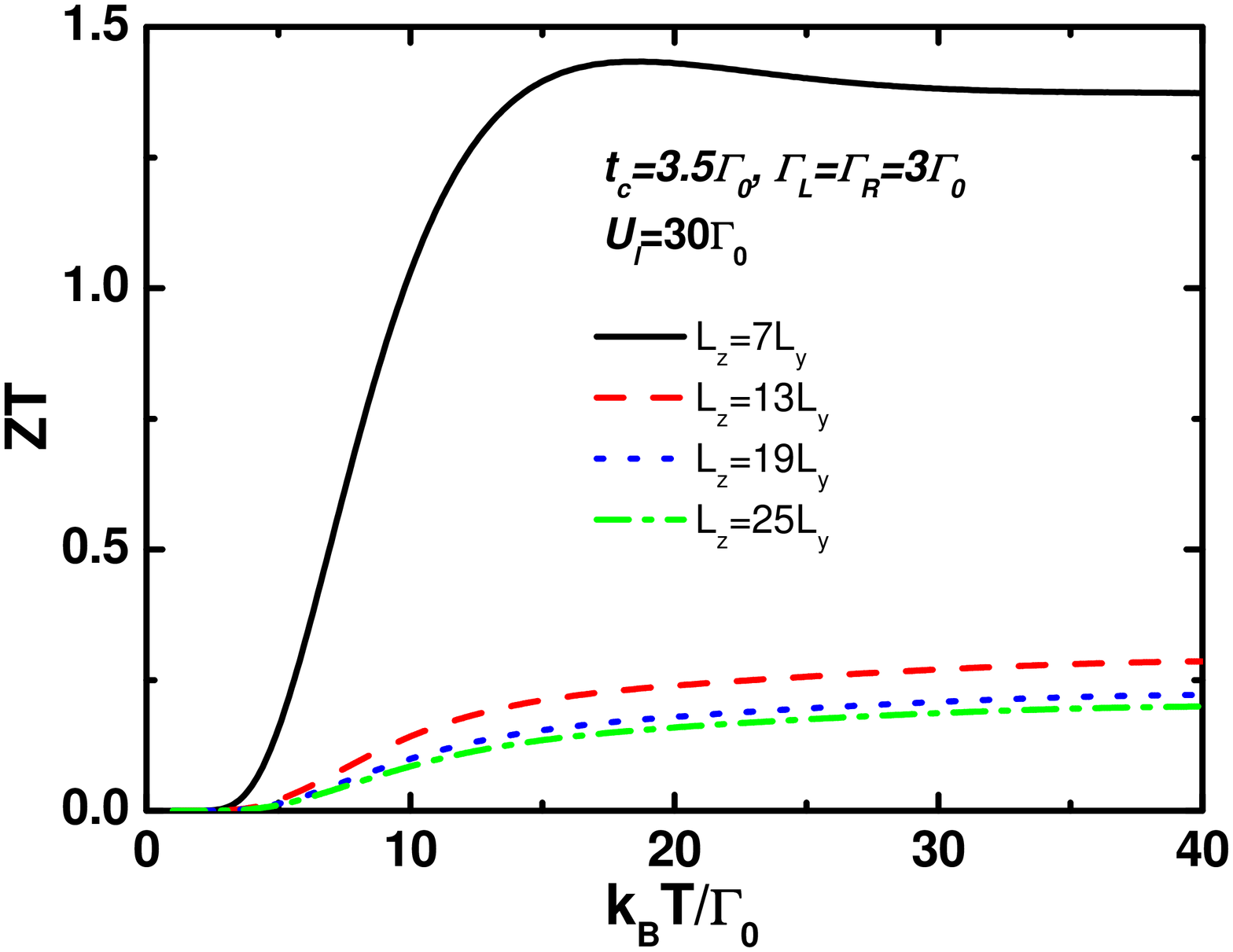}
\caption{$(ZT)$ as a function of temperature for different $L_z$
values. The physical parameters are $t_c=3.5\Gamma_0$ and
$\Gamma_L=\Gamma_R=3\Gamma_0$. Solid line ($L_z=7L_y$), dashed line
($L_z=13L_y$), dotted line ($L_z=19L_y$) and dot-dashed line
($L_z=25L_y$). Other physical parameters are the same as those of
Fig. 5(b).}
\end{figure}

\end{document}